\documentclass[5p]{elsarticle}
\usepackage[english]{babel}
\usepackage{graphicx}
\usepackage{amsmath, amssymb, upgreek}
\graphicspath{{Figs/}}
\DeclareGraphicsExtensions{.png}

\journal{Journal of Magnetism and Magnetic Materials}

\begin{document}

\begin{frontmatter}

\title{Uniform and Nonuniform Precession of a Nanoparticle with Finite Anisotropy in a Liquid: Opportunities and Limitations for Magnetic Fluid Hyperthermia}

\author{T.~V.~Lyutyy\corref{cor_auth}}
\cortext[cor_auth]{Corresponding author}
\ead{lyutyy@oeph.sumdu.edu.ua}

\author{O.~M.~Hryshko}
\author{M.~Yu.~Yakovenko}

\address{Sumy State University, 2 Rimsky-Korsakov Street, UA-40007 Sumy, Ukraine}

\begin{abstract}
We focus on an in-depth study of the forced dynamics of a ferromagnetic single-domain uniaxial nanoparticle placed in a viscous fluid and driven by an external rotating magnetic field. The process of conversion of magnetic and mechanical energies into heat is a physical basis for magnetic fluid hyperthermia that is very promising for cancer treatment. The dynamical approximation allows us to establish the limits of the heating rate and understand the logic of selection of the system parameters to optimize the therapy. Based on the developed analytical and numerical tools, we analyze from a single viewpoint the synchronous and asynchronous rotation of the nanoparticle or/and its magnetization in the following three cases. For the beginning, we actualize the features of the internal magnetic dynamics, when the nanoparticle body is supposed to be fixed. Then, we study the rotation of the whole nanoparticle, when its magnetization is supposed to be locked to the crystal lattice. And, finally, we realize the analysis of the coupled motion, when the internal magnetic dynamics is performed in the rotated nanoparticle body. In all these cases, we describe analytically the uniform mode, or synchronous rotation along with an external field, while the nonuniform mode, or asynchronous rotation, is investigated numerically.
\end{abstract}


\end{frontmatter}


\section{Introduction}
Ferrofluid \cite{Rosensweig1985Ferrohydrodynamics, 0038-5670-17-2-R02} applications in biotechnologies and medicine are the key of stable interest to these media in recent years. Here, one should note targeted drug delivery \cite{0022-3727-36-13-201, VEISEH2010284}, biosensors, macro-molecule and virus separation \cite{0022-3727-36-13-201, TIAN2016420, C6AY00721J}, magnetic fluid hyperthermia \cite{0022-3727-36-13-201, JORDAN1999413, 0957-4484-25-45-452001}. The latter is a promising method with principal advantages over conventional chemotherapy. First, hyperthermia is a local method and concentrates its action on the injured tissue. Second, hyperthermia operates with nontoxic matters and has comparatively lower side effects. Here, treatment is realized through the local heating provided due to the absorption of an external alternating field by ferromagnetic nanoparticles, which were injected and concentrated around the injured tissue. There are three mechanisms of conversion of the field energy into heat: 1) the rotation of the nanoparticle body in a viscous carrier, 2) the damped precession of the particle magnetization inside the nanoparticle, 3) the Eddy currents inside the particle, which are induced by an external field. The latter is negligibly small, and further we consider the first two cases.

Strictly speaking, magnetic dynamics and mechanical motion interfere with each other. However, in literature the simplified models are often used to describe the interaction of the nanoparticle with an alternating field. Supposing that the nanoparticle is immobilized, the problem of energy dissipation for the nanoparticle driven by a circularly polarized field is considered both analytically and numerically in \cite{0953-8984-21-12-124202, PhysRevB.91.054425}. In the mentioned investigations, based on the noise-free single-particle model, the exact expressions for the power loss were obtained in some specific cases, and the numerical simulation has been performed for the general case. The simulations of the nanoparticle ensemble, where only the magnetic damped precession produces the losses, are reported in \cite{PhysRevB.85.045435, PhysRevB.89.014403}. Despite the results obtained in the cited works have sufficient scientific relevance, they do not exhaust the hyperthermia problem, because the motion of the nanoparticle bodies in most cases has considerable influence on losses.

In this regard, another approximation, which should be considered, is the so-called the rigid dipole (or frozen magnetic moment) model. Within this, the nanoparticle magnetization is assumed to be fixed to its anisotropy axis, and the nanoparticle performs the mechanical motion only. Using the rigid dipole model, the response to an external alternating field and the power losses were treated in terms of the complex magnetic susceptibility \cite{PhysRevE.63.011504, PhysRevE.83.021401, 0953-8984-15-23-313, SotoAquino201546}. The detailed microscopic consideration of the stochastic dynamics of such nanoparticles was given in \cite{0953-8984-15-23-313, PhysRevE.92.042312}, and the influence of the dipole interaction on the power loss was studied in detail in \cite{PhysRevE.97.052611}. Despite the rigid dipole model is partially valid for the system parameters, which are suitable for hyperthermia, even small deviations of the magnetization from the easy axis can lead to the significant changes in the dissipation process.

The coupled dynamics of the nanoparticle body and its magnetization are complex and cannot be described by a simple superposition of the above mentioned types of motion. The model equations were firstly written in \cite{Cebers1975}, but the discussion of their explicit form has been continued until now \cite{Mamiya2011,doi:10.1063/1.4737126,doi:10.1063/1.4937919,Usov2015339}. Special attention deserves also the attempt to describe the energy absorption during the forced coupled motion presented in \cite{0022-3727-39-22-002}. There, the power loss was obtained on the basis of the Lagrangian equation in the noise-free and single-particle approximation. After this, the progress in the description of the energy dissipation of a viscously coupled nanoparticle with finite anisotropy driven by an alternating field was achieved in \cite{PhysRevB.95.134447, doi:10.21272/jnep.8(4(2)).04086, Lyutyy201887}. The further development of this model assumes the accounting of thermal noise firstly. In this regard, several important results were obtained in \cite{PhysRevB.95.104430, 0031-9155-63-3-035004}, but the role of thermal fluctuations in the energy absorbtion has not yet found.

The analysis performed suggests that even in the deterministic, single-particle case, some issues remain unclear that complicates  further proceeding. Therefore, in the present study we continue to develop the methodological scheme, which is realized in \cite{Lyutyy201887}. To understand better the role of both energy dissipation channels, we consider consequently and on common footing the driven magnetic dynamics in the fixed nanoparticle, the mechanical rotation of the nanoparticle with the fixed magnetization, and, finally, the coupled motion of the nanoparticle with finite anisotropy. We assume that the rotating field acts, and there are two types of motion, i.e. the uniform and nonuniform precession. The first type is treated analytically, while for the latter the numerical description is demanded.

\section{Model and Basic Equations}
We consider a uniform spherical single-domain ferromagnetic nanoparticle of radius $R$, magnetization $\mathbf{M}$ ($\mathbf{|M|}= M = \mathrm{const}$), and density $\rho$. This particle performs the spherical motion (or motion with the fixed center of mass) with respect to a fluid of viscosity $\eta$. Then, we assume that the nanoparticle is driven by the external circularly polarized field
\begin{equation}
    \mathbf{H}(t) = \mathbf{e}_{x} H\cos(\Omega t) + \mathbf{e}_{y} \varrho H\sin(\Omega t),
    \label{eq:def_h}
\end{equation}
where $\mathbf{e}_x$, $\mathbf{e}_y$, $\mathbf{e}_z$ are the unit vectors of the Cartesian framework, $H$ is the field amplitude, $\Omega$ is the field frequency, $t$ is the time, and $\varrho$ is the factor, which determines the polarization type ($\varrho=\pm1$).

The magnetic energy of the nanoparticle is given by
\begin{equation}
    W = -\dfrac{H_{a}V}{2M} (\mathbf{M}\mathbf{n})^{2} - V \mathbf{M}\mathbf{H}(t) - V H_z \mathbf{M} \mathbf{e}_{z},
    \label{eq:def_W}
\end{equation}
where $H_{a}$ is the magnitude of the uniaxial anisotropy, $\mathbf{n}$ is the unit vector defining the anisotropy axis direction, $H_z$ is the magnitude of the constant field. Taking into account the action on the nanoparticle magnetization of its crystal lattice, the effective magnetic field acting on the nanoparticle can be written as
\begin{equation}
    \mathbf{H}_{\mathrm{eff}} = -V^{-1}\frac{\partial W}{\partial \mathbf{M}} =
    H_{a}M^{-1} (\mathbf{M}\mathbf{n})\mathbf{n} + \mathbf{H}(t) + \mathbf{e}_{z} H_z.
    \label{eq:H_eff}
\end{equation}
The dynamics of the nanoparticle leads to the dissipation of its energy $W$. In accordance with \cite{PhysRevB.91.054425}, the power loss is determined as $Q = \lim_{\tau \to \infty} (1/\tau) \int_{0}^{\tau} dt q$, where $q = -dW/dt$ is the instantaneous power loss. As follows from Eq.~(\ref{eq:def_W}), $q = V \mathbf{H}_{\mathrm{eff}}\cdot d\mathbf{M}/dt$, and the reduced power loss $\widetilde{Q} = Q/(H_{a} V M \Omega_0)$ ($\Omega_{0}$ is some characteristic frequency, which is determined on the model approach) is written in the form
\begin{equation}
    \widetilde{Q} = \lim_{\widetilde{\tau} \to \infty}
    \frac{1} {\widetilde{\tau}} \int_{0}^{\widetilde{\tau}} d\tilde{t}\,
    \mathbf{h}_{\mathrm{eff}} \cdot \dot{\mathbf{m}}.
    \label{eq:def_Q}
\end{equation}
Here, $\mathbf{h}_{\mathrm{eff}} = \mathbf{H}_{\mathrm{eff}} /H_{a}$ is the reduced effective field, $\mathbf{m} = \mathbf{M}/M$ is the unit vector, which represents the direction of the nanoparticle magnetization, $\tilde{t} = \Omega_{0} t$ is the reduced time, $\widetilde{\tau} = \Omega_{0} \tau$. It is important to highlight that Eq.~(\ref{eq:def_Q}) is suitable for both the analytical and numerical further treatment of the energy dissipation problem.

As stated above, for the analytical description there are three approaches to the nanoparticle dynamics: 1) the nanoparticle body is fixed, and the magnetic moment can rotate about the anisotropy axis; 2) the magnetic moment is fixed within the viscously rotated nanoparticle; 3) both the nanoparticle body and its magnetic moment perform the rotation. Let us describe all these approaches in detail.

\subsection{Internal magnetic dynamics}
In the case of high frequencies, large enough particles and/or carrier viscosities, and not so large anisotropy, the nanoparticle body motion is negligible. Here, only the magnetic moment dynamics should be considered. To these purposes, the well-known Landau-Lifshitz-Gilbert equation can be utilized
\begin{equation}
    \mathbf{\dot{M}}=-\gamma \mathbf{M}\times \mathbf{H}_{\mathrm{eff}} +\alpha {M}^{-1}\mathbf{M}\times  \mathbf{\dot{M}}.
    \label{eq:Main_Eq_FP_m}
\end{equation}
Here, $\gamma (>0)$ is the gyromagnetic ratio, $\alpha(>0)$ is the dimensionless damping parameter.

In the dimensionless form, Eq.~(\ref{eq:Main_Eq_FP_m}) can be rewritten as
\begin{equation}
    \mathbf{\dot{m}}=-\Omega_r \mathbf{m}\times \mathbf{h}_{\mathrm{eff}} +\alpha \mathbf{m}\times \mathbf{\dot{m}}.
    \label{eq:Red_Eq_FP_m}
\end{equation}
Using recursive substitution and taking into account the properties of the vector product, it is easy to show that Eq.~(\ref{eq:Red_Eq_FP_m}) corresponds to
\begin{equation}
    (1 + \alpha^2)\mathbf{\dot{m}}=-\Omega_r \mathbf{m}\times \mathbf{h}_{\mathrm{eff}} -\alpha \Omega_r \mathbf{m}\times
    \mathbf{m}\times \mathbf{h}_{\mathrm{eff}},
    \label{eq:Red_Eq_FP_m_1}
\end{equation}
which is more convenient for the numerical treatment. After standard transformations and accounting Eq.~(\ref{eq:def_h}), one can write the set of scalar equations with respect to the polar $\vartheta$ and azimuthal $\varphi$ angles of the vector $\mathbf{m}$
\begin{eqnarray}
    (1+\alpha^{2})\Omega_{r}^{-1}\dot{\vartheta} \!\!\!\!&=&\!\!\!\! \alpha h \cos\vartheta f\!+\!h f_{\varphi}\!-\!\alpha \sin \vartheta(\cos\vartheta\!+\!h_z),
    \nonumber\\ [6pt]
    (1+\alpha^{2})\Omega_{r}^{-1}\dot{\varphi} \!\!\!\!&=&\!\!\!\! \alpha h \csc\vartheta f_{\varphi}\!-\!h\cot\vartheta f\!+\!
     \cos\vartheta\!+\!h_z, \nonumber\\
\label{eq:FixPart_angles}
\end{eqnarray}
where $h = H/H_{a}$, $h_{z} = H_{z}/H_{a}$,
\begin{equation}
    f = \cos\varphi \cos(\Omega t) + \varrho\sin\varphi \sin(\Omega t),
    \label{eq:def_f}
\end{equation}
and $f_{\varphi} = \partial f/ \partial \varphi$.

\subsection{Motion of the nanoparticle body}
In the case of strong anisotropy or weak coupling with the environment, the internal magnetic dynamics can be negligible. And here, the nanoparticle dynamics is described by the rigid dipole model, when the magnetization is supposed to be fixed to the anisotropy axis. This model is introduced in \cite{doi:10.1063/1.1656014} and has been successfully used up to now. The main peculiarity of the analytical description is the presence of two vector equations. The first equation, in fact, is the condition of the rigid body rotation, and the second one is the second Newton's law for the rotational motion
\begin{equation}
\begin{array}{lcl}
    \dot{\mathbf{n}}= \boldsymbol{\upomega}\times \mathbf{n}, \\
    [2pt]
     J \dot{\boldsymbol{\upomega}} = VM \mathbf{n}\times \mathbf{H}- 6\eta V\boldsymbol{\upomega}.\\
    \label{eq:Main_Eq_FM_n}
\end{array}
\end{equation}
Here, $\boldsymbol{\upomega}$ is the nanoparticle angular velocity, $J(= 8\pi \rho R^{5}/15)$ is the nanoparticle moment of inertia, $V$ is the nanoparticle volume, and dots over symbols represent derivatives with respect to time. When the inertia momentum is too small and can be neglected, Eqs.~(\ref{eq:Main_Eq_FM_n}) are transformed into a simple form
\begin{equation}
    \dot{\mathbf{n}}= - \Omega_{cr}\mathbf{n}\times\left(\mathbf{n}\times \mathbf{h}\right),
    \label{eq:Main_Eq_FM_n1}
\end{equation}
where $\Omega_{cr}=M H_{a}/(6 \eta)$ is the characteristic frequency of the uniform mechanical rotation. After standard transformations and accounting Eq.~(\ref{eq:def_h}), one can write the set of scalar equations with respect to the polar $\theta$ and azimuthal $\phi$ angles of the vector $\mathbf{n}$
\begin{equation}
    \begin{array}{ll}
    \displaystyle \Omega_{cr}^{-1}\, \dot{\theta} = h\cos\theta \cos(\varrho \Omega t-\phi) - h_{z} \sin\theta,
    \\ [6pt]
    \displaystyle \Omega_{cr}^{-1}\, \dot{\phi} = h \sin^{-1}\theta\sin(\varrho \Omega t-\phi).
    \end{array}
    \label{eq:Main_Eq_spher_cp}
\end{equation}

\subsection{The coupled dynamics of the body and magnetic moment of the nanoparticle}
As shown in detail in \cite{doi:10.1063/1.4937919}, the coupled magnetic dynamics and the mechanical motion cannot be described by a simple superposition of these two types of motion because of the significant changes in the basic equations. Ultimately, it was stated that the coupled dynamics obeys the following pair of coupled equations:
\begin{equation}
\begin{array}{lcl}
    \dot{\mathbf{n}}= \boldsymbol{\upomega}\times \mathbf{n}, \\
    [2pt]
    J \dot{\boldsymbol{\upomega}}=\gamma^{-1}V\dot{\mathbf{M}} + V\mathbf{M}\times \mathbf{H}- 6\eta V\boldsymbol{\upomega},\\
    \label{eq:Main_Eq_FA_n}
\end{array}
\end{equation}
\begin{equation}
    \mathbf{\dot{M}}=-\gamma  \mathbf{M}\times \mathbf{H}_{eff} +\alpha {M}^{-1}\left(\mathbf{M}\times  \mathbf{\dot{M}}-\boldsymbol{\upomega}\times \mathbf{M} \right).
    \label{eq:Main_Eq_FA_m}
\end{equation}
In the case when the inertia term in (\ref{eq:Main_Eq_FA_n}) is negligible, this equation can be transformed into the more convenient form. Then, we transform the equation for the internal magnetic dynamics (\ref{eq:Main_Eq_FA_m}) in order to separate the terms containing the time derivatives. As a result, we obtain
\begin{equation}
\begin{array}{lcl}
    \Omega_{cr}^{-1}\,\mathbf{\dot{n}}=\mathbf{\dot{m}}\times \mathbf{n}/\Omega_{r}+\left(\mathbf{m}\times \mathbf{h} \right)\times \mathbf{n},\\
    [4pt]
    (1+\alpha^{2}_{1}) \Omega_{r1}^{-1} \mathbf{\dot{m}}=- \mathbf{m}\times \mathbf{h}_{eff}^{1}-\alpha_{1} \mathbf{m}\times\mathbf{m}\times \mathbf{h}_{eff}^{1},
    \label{eq:Red_Eq_FA}
\end{array}
\end{equation}
where $\beta =\alpha M/6\gamma \eta$, $\Omega_{r1} = \Omega_{r}/(1+\beta)$, $\alpha_{1} = \alpha/(1+\beta)$,
\begin{equation}
    \mathbf{h}_{eff}^{1}=\left(\mathbf{e}_{x} h\cos{\Omega t} +\mathbf{e}_{y} \varrho h \sin{\Omega t} \right)\left(1+\beta  \right)+\left(\mathbf{mn} \right)\mathbf{n}.
    \label{eq:h_eff_1}
\end{equation}
After standard transformations and accounting Eq.~(\ref{eq:def_h}), we can write the set of scalar equations with respect to the polar $\theta$ and azimuthal $\phi$ angles of the vector $\mathbf{n}$, as well as to the polar $\vartheta$ and azimuthal $\varphi$ angles of the vector $\mathbf{m}$
\begin{equation}
    \begin{array}{rll}
    \displaystyle (1+\alpha^{2}_{1}) \Omega_{r1}^{-1}\, \dot{\vartheta} \!&=& f_1 + \alpha_1 f_2,
    \\ [6pt]
    \displaystyle (1+\alpha^{2}_{1}) \Omega_{r1}^{-1}\, \dot{\varphi} \!&=& \sin^{-1}\vartheta(\alpha_1 f_1 - f_2),
    \\ [6pt]
    \displaystyle \Omega_{r1}^{-1}\, \dot{\theta} \!&=& \beta \alpha^{-1} (\omega_{y} \cos \phi - \omega_{x} \sin \phi),
    \\ [6pt]
    \displaystyle \Omega_{r1}^{-1}\, \dot{\phi} \!&=& \beta \alpha^{-1} \big[\omega_{z} -\cot\theta (\omega_{y} \sin\phi\\
    && +\, \omega_{x} \cos \phi)\big],
    \end{array}
    \label{eq:fin_anis_base_eq}
\end{equation}
where
\begin{equation}
    \begin{array}{rll}
    \displaystyle f_1 \!&=& \big[ h(1 + \beta)\sin(\varrho \Omega t-\phi) - F \sin \theta \sin(\varphi - \phi)\big],
    \\ [6pt]
    \displaystyle f_2 \!&=& \cos\vartheta \big[ h(1 + \beta)\cos(\varrho \Omega t-\phi)  \\
     &&+\, F \sin \theta \cos(\varphi - \phi)\big] - \sin\vartheta \big[(1 + \beta)h_z\\
     &&+\, F\cos\theta \big],
    \\ [6pt]
    \displaystyle F \!&=& \cos \theta \cos\vartheta + \cos(\varphi - \phi) \sin \theta \sin\vartheta \, (= \mathbf{m}\mathbf{n})
    \\ [6pt]
    \displaystyle\omega_{x} \!&=&   \dot{\vartheta} \cos\vartheta \cos\varphi  +
    \dot{\varphi}\sin\vartheta \sin \varphi  \\
    && - \, (1 + \beta) \big[ h_z \sin\vartheta \sin\varphi + h \cos\vartheta \cos(\Omega t) \big],
    \\ [6pt]
    \displaystyle\omega_{y} \!&=&  \dot{\vartheta} \cos\vartheta \sin\varphi  +
    \dot{\varphi}\sin\vartheta \cos \varphi \\
    && - \, (1 + \beta) \big[ h_z \sin\vartheta \cos\varphi + h \cos\vartheta \sin(\varrho\Omega t) \big],
    \\ [6pt]
    \displaystyle\omega_{z} \!&=& (1 + \beta) h \sin(\varrho\Omega t - \varphi)\sin \vartheta - \dot{\vartheta} \sin \vartheta.
    \end{array}
    \label{eq:fin_anis_base_eq_designat}
\end{equation}
We want to underline here that the system Eqs.~(\ref{eq:fin_anis_base_eq}) along with designations Eqs.~(\ref{eq:fin_anis_base_eq_designat}) are appropriate for further numerical treatment.

Therefore, the model equations derived above allow us to perform the investigation of the precessional motion of the nanoparticle induced by the external circularly polarized field. The approach used neglects thermal fluctuations. Its validity is discussed in \cite{PhysRevB.91.054425, Lyutyy201887}. The forced stochastic motion in the simplified cases of the rigidly fixed nanoparticle and the rigid dipole are considered in \cite{PhysRevLett.97.227202, PhysRevB.84.174410} and in \cite{PhysRevE.92.042312, PhysRevE.97.052611}, respectively. The stochastic motion in the case of the coupled magnetic dynamics and the mechanical rotation is not completely studied yet. Some issues are discussed in \cite{PhysRevB.95.104430, 0031-9155-63-3-035004}.

\section{Results and Discussion}
\subsection{Internal magnetic dynamics}
If the nanoparticle is supposed to be immobilized, there are two modes of the steady-state dynamics of $\mathbf{m}$ under the action of the field of type Eq.~(\ref{eq:def_h}) \cite{PhysRevLett.86.724, 0953-8984-21-39-396002, DENISOV20101360}. The first mode is the uniform rotation, which is performed synchronously with the external field. The second one is the nonuniform rotation, when the period of $\mathbf{m}$ does not coincide with the period of $\mathbf{H}(t)$. From the analytical viewpoint, the uniform mode is characterized by the constant precession and lag angles, $\vartheta_1$ and $\varphi_1$, where $\varphi_1 = \varphi-\varrho \Omega t$. As follows from Eqs.~(\ref{eq:FixPart_angles}), the precession angle satisfies the equation\cite{PhysRevLett.86.724, PhysRevLett.97.227202}
\begin{equation}
    h^2 = \frac{1 - \cos^{2}\vartheta_1}{\cos^{2}\vartheta_1}
    \bigg[\bigg(\cos\vartheta_1 + h_z -\frac{\varrho\widetilde{\Omega}} {1 + \alpha^2}\bigg)^2
    + \bigg(\frac{\alpha\widetilde{\Omega}\cos\vartheta_1}{1 + \alpha^2}\bigg)^2\bigg],\\[4pt]
    \label{eq:FP_Theta}
\end{equation}
and the lag angle is connected to the precession one as
\begin{equation}
    \sin\varphi_1 = -\varrho\frac{\alpha\widetilde{\Omega}} {h(1 + \alpha^2)}\sin\vartheta_1.
    \label{eq:FP_Phi}
\end{equation}
Here $\Omega_0 = \Omega_r$, $\widetilde{\Omega} = \Omega/\Omega_r$. After integration by parts of Eq.~(\ref{eq:def_Q}), we obtain the general expression for the reduced power loss in the case of the periodic mode
\begin{equation}
    \widetilde{Q} = \alpha \frac{\widetilde{\Omega}^{2}}{(1 + \alpha^2)^2} \sin^{2}\vartheta_1.
    \label{eq:Q_FP}
\end{equation}
In the nonuniform mode, the polar angle $\vartheta$ of the vector $\mathbf{m}$ varies periodically in time with a period, which does not coincide with the field one. The similar oscillations are demonstrated by the azimuthal angle $\varphi$ together with the linear growth in time. This dynamics is accompanied by the power losses, which can be investigated only in the numerical way. The difference scheme for the numerical calculus of the power loss here is written as
\begin{eqnarray}
    \displaystyle \widetilde{Q} \!\!\!&=&\!\!\! \frac{1} {N} \sum^{N}_{i = 1}
    \bigg[h_{xi}(\cos\vartheta_i\cos\varphi_i \Delta\vartheta_i - \sin\vartheta_i\sin\varphi_i \Delta\phi_i)\nonumber
    \\
    \displaystyle \!&& +\,h_{yi} (\cos\vartheta_i\sin\varphi_i \Delta\theta_i + \sin\vartheta_i\cos\varphi_i \Delta\varphi_i) \nonumber
    \\
    \displaystyle \!&& -\,h_{zi}\sin\vartheta_i\Delta\vartheta_i\bigg],
    \label{eq:Q_FM_Num}
\end{eqnarray}
where $N = \widetilde{\tau}/\Delta\tilde{t}$ ($\widetilde{\tau} = \tau \Omega_{r}$ and is chosen as $10^5$ in the simulation) is the number of time steps on the external field period, $\Delta \tilde{t}(\ll 1/\widetilde{\Omega})$ is the value of the time step within the numerical calculation procedure, $\vartheta_i = \vartheta(\tilde{t}_i)$, $\varphi_i = \varphi(\tilde{t}_i)$, $\Delta\vartheta_i = \frac{\partial \vartheta(\tilde{t}_i)}{\partial \tilde{t}}\Delta\tilde{t}$, $\Delta\varphi_i = \frac{\partial \varphi(\tilde{t}_i)}{\partial\tilde{t}}\Delta\tilde{t}$, $h_{xi} = h\cos(\varrho \widetilde{\Omega} \tilde{t}_i)$, $h_{yi} = h\sin(\varrho \widetilde{\Omega}\tilde{t}_i)$, $h_{zi} = h_z + \cos\vartheta_i$.

\begin{figure}
    \centering
    \includegraphics [width=1.0\linewidth] {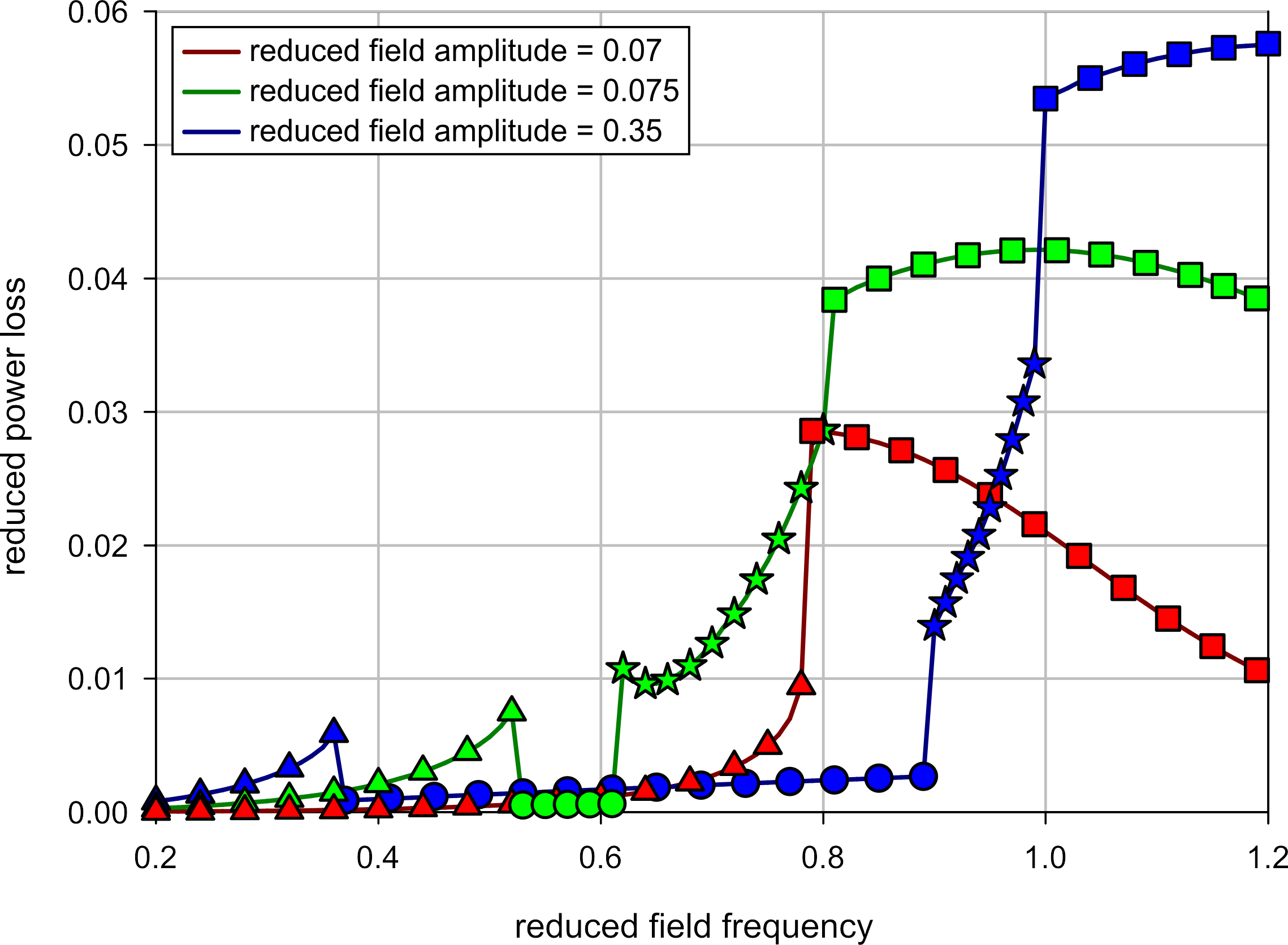}
    \caption {\label{fig:FP_Q_vs_w} (Color online) Model of the fixed nanoparticle: the most typical dependencies of the power loss on the field frequencies for different field amplitudes. The values of the system parameters are the following: $\alpha = 0.1$, $\varrho = +1$, $h_z = 0$. Triangle markers designate the uniform precession in the "up state"; circle markers designate the uniform precession after the magnetization switching to the "down state"; stars markers designate the nonuniform precession; square markers designate the uniform precession in the "up state" again.}
\end{figure}

The results of the series of simulations are illustrated in Fig.~\ref{fig:FP_Q_vs_w}. For the uniform mode, these results are in excellent agreement with those obtained from Eq.~(\ref{eq:Q_FP}). The sharp changes of $\widetilde{Q}$ are associated with the changes in the precession modes that is discussed in detail in \cite{PhysRevB.91.054425, 0953-8984-21-39-396002, DENISOV20101360}. When the field amplitude is considered to be constant, the most complicated case, which corresponds to the frequencies near the resonant one, is realized in the following way (see the curves for $h=0.21$ and $h=0.35$). For beginning, the power loss increases with the field frequency within the uniform mode, see the curves fractures with the triangle markers. Then, an abrupt increase in $\widetilde{Q}$ is caused by the reorientation or switching to the "down state", see the curves fractures with the circle markers. After that, the nonuniform mode starts to be generated that is testified by a sharp increase in the $\widetilde{Q}(\widetilde{\Omega})$ curve, see star markers. It is important, the condition $\vartheta<\pi/2$ holds predominantly. From the view point of energy minimizing, this mode is generated in order to reduce the losses that is clear from the figure. Finally, a further sharp increase in $\widetilde{Q}$ is the consequence of switching to the uniform mode again, see curves fractures with the squares.

\subsection{Motion of the nanoparticle body}
If the nanoparticle magnetic moment is fixed inside, the uniform and nonuniform precession modes can be also realized. The first of them is the natural solution of Eqs.~(\ref{eq:Main_Eq_spher_cp}). This mode is characterized by the constant lag angle $\phi_1 = \phi -\varrho \Omega t$ and the constant angle of the precession cone $\theta_1$. Substituting these solutions into Eqs.~(\ref{eq:Main_Eq_spher_cp}), we derive the system of algebraic equations for the calculation of $\phi_1$ and $\theta_1$
\begin{equation}
\begin{array}{lcl}
    \cos \theta_1\left(\widetilde{\Omega}^2\alpha\sin\theta_1 + h\cos\phi_1 \right) = h_{z}\sin\theta_1, \\
    \widetilde{\Omega}\sin\theta_1 =h\sin\phi_1.\\
    \label{eq:Main_Eq_spher_cp_sol}
\end{array}
\end{equation}
Here, $\Omega_0 = \Omega_{cr}$ and $\widetilde{\Omega} = \Omega/\Omega_{cr}$. The average value of the power loss can be found easily in this case. The straightforward calculations using Eqs.~(\ref{eq:Main_Eq_spher_cp_sol}) and Eq.~(\ref{eq:def_Q}) yield
\begin{equation}
    \widetilde{Q} = \widetilde{\Omega}^{2}\sin^{2}\theta_1.
    \label{eq:Q_RD}
\end{equation}
Two remarks are relevant here. First, Eq.~(\ref{eq:Q_RD}) for a small angle of the precession cone coincides with the results obtained by Xi \cite{0022-3727-39-22-002} in the linear approximation. And, second, when the static field is absent ($h_{z} = 0$), the relationships $\theta_1 =\pi/2$, $\sin \phi_1 = \widetilde{\Omega}/h$, and $\widetilde{Q} = \widetilde{\Omega}^{2}$ are valid.

To describe the power loss behavior in the whole range of parameters and visualize the data, the numerical simulation is also demanded here. The difference scheme for the numerical calculus of the power loss is written as
\begin{eqnarray}
    \widetilde{Q} \!\!\!&=&\!\!\! \frac{1} {N} \sum^{N}_{i = 1}
    \bigg[h_{xi}(\cos\theta_i\cos\phi_i \Delta\theta_i - \sin\theta_i\sin\phi_i \Delta\phi_i)\nonumber
    \\
    && +\,h_{yi}(\cos\theta_i\sin\phi_i \Delta\theta_i + \sin\theta_i\cos\phi_i \Delta\phi_i) \nonumber
    \\
    && -\,h_z\sin\theta_i\Delta\theta_i\bigg],
    \label{eq:Q_FM_Num}
\end{eqnarray}
where $N = \widetilde{\tau}/\Delta\tilde{t}$ ($\widetilde{\tau} = \tau \Omega_{cr}$ and is chosen as $10^5$ in the simulation)  is the number of time steps on the external field period, $\Delta \tilde{t}(\ll 1/\widetilde{\Omega})$ is the value of the time step within the numerical calculation procedure, $\theta_i = \theta(\tilde{t}_i)$, $\phi_i = \phi(\tilde{t}_i)$, $\Delta\theta_i = \frac{\partial \theta(\tilde{t_i})}{\partial \tilde{t}}\Delta\tilde{t}$, $\Delta\phi_i = \frac{\partial \phi(\tilde{t}_i)}{\partial \tilde{t}}\Delta\tilde{t}$, $h_{xi} = h\cos(\varrho \widetilde{\Omega} \tilde{t}_i)$, $h_{yi} = h\sin(\varrho \widetilde{\Omega}\tilde{t}_i)$.

\begin{figure}
    \centering
    \includegraphics [width=1.0\linewidth] {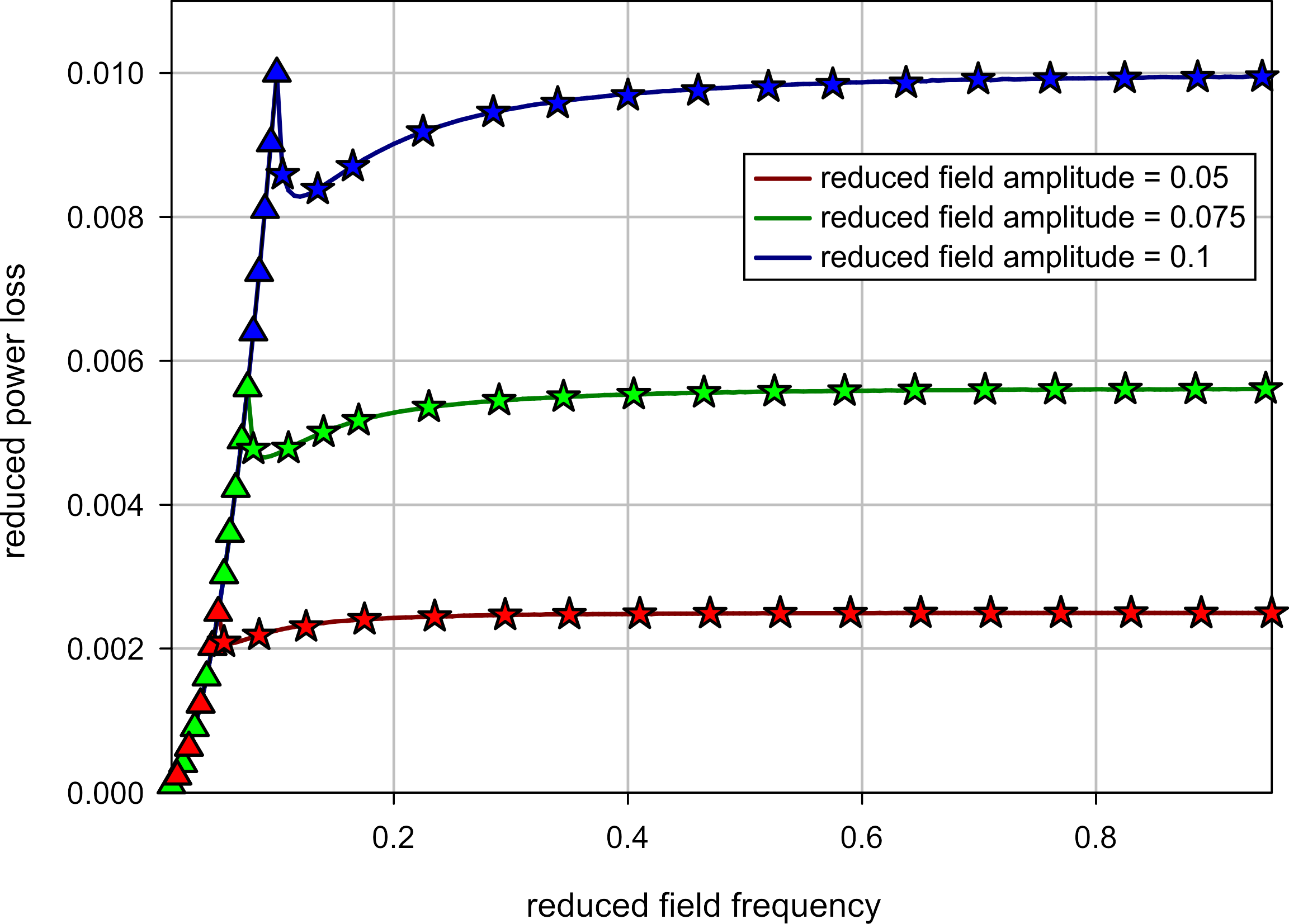}
        \caption {\label{fig:FM_Q_vs_w} (Color online) Model of the fixed magnetization: the most typical dependencies of the power loss on the field frequencies for different field amplitudes. The values of the system parameters are the following: $\eta = 0.05 P$, $\varrho = +1$, $h_z = 0$, and the initial condition $\theta_0 = 0.01$. Triangle markers designate the uniform precession; star markers designate the nonuniform precession.}
\end{figure}

As follows from the analytical results discussed above, when $h > \widetilde{\Omega}$, the nanoparticle is rotated uniformly, and all contributions into the power loss are due to this rotation. This is confirmed by the series of simulations, the results of which are shown in Fig.~\ref{fig:FM_Q_vs_w}, see the triangle markers. At the same time, when $h < \widetilde{\Omega}$ and $h \sim \widetilde{\Omega}$, the dynamics becomes nonuniform, see the star markers. Similar to the previous case, here $\mathbf{n}$ performs the rotation simultaneously with the oscillations of a larger period. Since the nonuniform precession is characterized by smaller instantaneous angular velocity of the nanoparticle, a decrease in the power loss is observed. It is expressed in a pronounced drop of $\widetilde{Q}(\widetilde{\Omega})$ for the fixed amplitude $h$ (see Fig.~\ref{fig:FM_Q_vs_w}). There are two features, which should be underlined in this regard. First, while the field frequency grows, the average angular velocity tends to zero, the oscillation frequency tends to $\widetilde{\Omega}$, and the oscillation amplitude tends to the saturated values predicted by Eq.~(34) in \cite{Lyutyy201887}. Second, the resulting power loss in the nonuniform mode depends on the initial position of the nanoparticle.

\subsection{The coupled dynamics of the body and magnetic moment of the nanoparticle}
\begin{figure}
    \centering
    \includegraphics[width=0.6\linewidth]{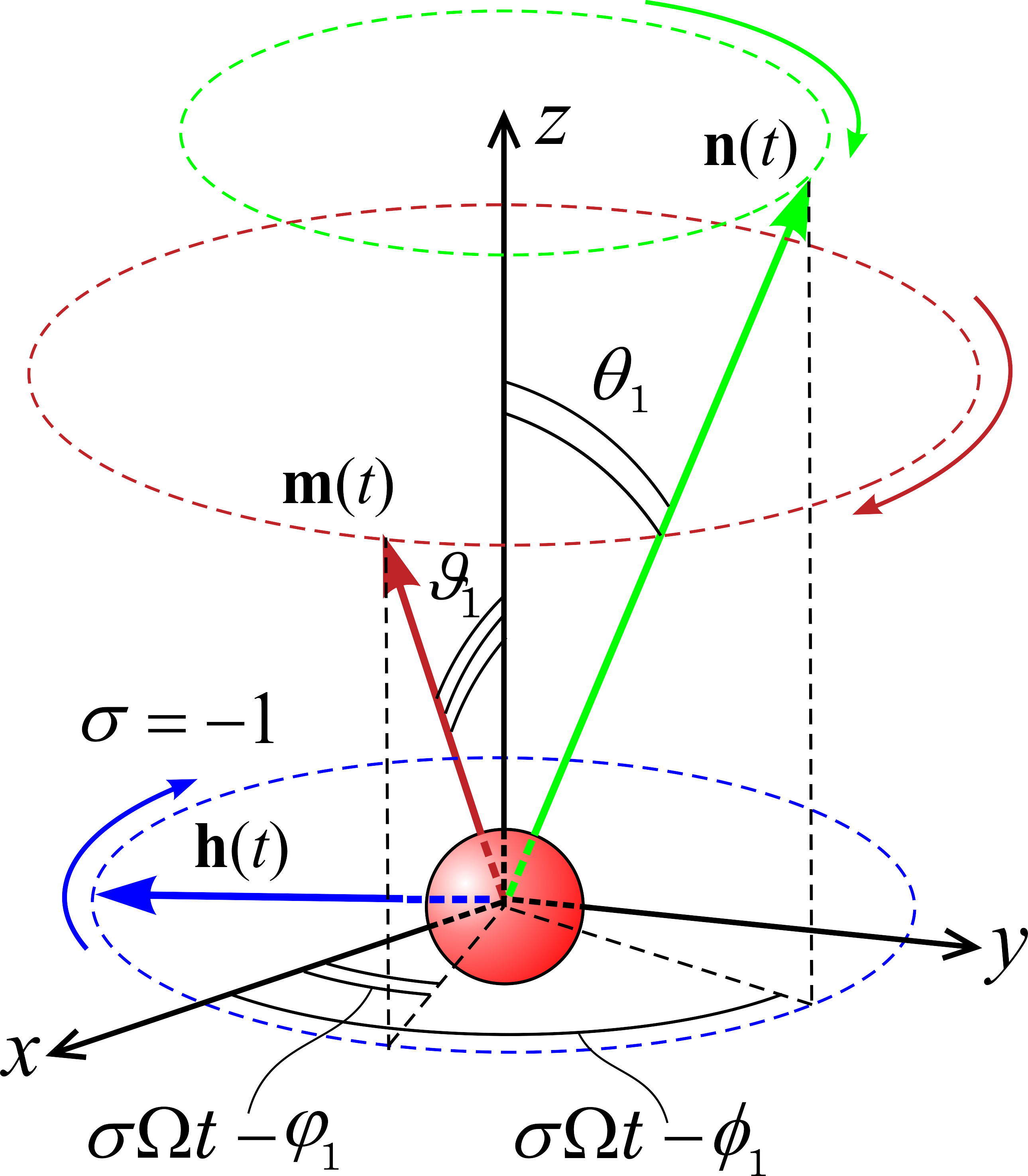}
    \caption{Schematic representation of the behaviour of the vectors $\mathbf{n}$, $\mathbf{m}$ and the coordinate systems used for the precession mode description}
    \label{fig:model_FA}
\end{figure}

In the case of synchronous precession of the vectors $\mathbf{m}$ and $\mathbf{n}$ with the external circularly polarized field (see Fig.~\ref{fig:model_FA}), the stationary solution of the set of equations (\ref{eq:Main_Eq_FA_n}), (\ref{eq:Main_Eq_FA_m}) can be obtained in the form $\varphi = \varrho \Omega t -\varphi_1$,  $\vartheta = \vartheta_1$,  $\phi = \varrho \Omega t -\phi_1$,  $\theta = \theta_1$. To find the unknown constants $\varphi_1$, $\phi_1$ and $\vartheta_1$, $\theta_1$, we used the condition of absence of the magnetic moment motion with respect to the nanoparticle crystal lattice
\begin{equation}
       \dot{\mathbf{m}} - \boldsymbol{\upomega}\times\mathbf{m}=0.
       \label{eq:rot_cond}
\end{equation}
After substitution of Eq.~(\ref{eq:rot_cond}) into the second equation of Eqs.~(\ref{eq:Main_Eq_FA_n}) and neglecting the inertia term, we derive
\begin{equation}
    0 = \Omega_{cr}(\mathbf{m}\times\mathbf{n}) (\mathbf{m}\mathbf{n}) + \boldsymbol{\upomega}.\\
    \label{eq:Main_Eq_rot}
\end{equation}
Then, let us introduce the double-primed coordinate system $x'' y'' z''$, which is rotated along with the external field as follows from Fig.~\ref{fig:model_FA}. In this new framework, the angular velocity has a very simple form
\begin{equation}
        \boldsymbol{\upomega}'' = \left( - \varrho \Omega \sin \theta_1 , 0, 0\right).
        \label{eq:angilar_double prime}
\end{equation}
Since Eq.~(\ref{eq:Main_Eq_FA_m}) cannot be easily represented in the double-primed system, we need to write the explicit form of all the vectors in the laboratory coordinate system. To perform the necessary transformations, we should use the rotation matrix
\begin{equation}
    {\textbf{C}^{-1}} = \left(
        \begin{array}{lcr}
           \cos\theta_1 \cos\Phi_1 & -\sin\Phi_1 & \sin\theta_1 \cos\Phi_1 \\
           \cos\theta_1 \sin\Phi_1 &  \cos\Phi_1 & \sin\theta_1 \sin\Phi_1 \\
           - \sin\theta_1 & 0 & \cos\theta_1 \\
        \end{array}
        \right),
        \label{eq:C_minus}
\end{equation}
where $\Phi_1 = \varrho \Omega t - \phi_1$. Let us introduce the designation
\begin{equation}
    F_1 = \sin\theta_1 \sin\vartheta_1 \cos(\phi_1 - \varphi_1) + \cos\theta_1 \cos\vartheta_1
    \label{eq:F}
\end{equation}
and represent the vectors $\boldsymbol{\upomega}$, $\mathbf{m}$, and $\mathbf{n}$ in the laboratory system using the rotation matrix Eq.~(\ref{eq:C_minus}). This allows us to straightforwardly obtain the set of algebraical equations
    \begin{eqnarray}
         -\sin\vartheta_1 \sin\varphi_1\bigg(\dfrac{\varrho \Omega}{\Omega_{r}} - h_{z}\bigg)\!\!\!\!&=&\!\!\!\!   \sin\theta_1 \cos\theta_1  \cos\phi_1 \dfrac{\varrho \Omega}{\Omega_{cr}}, \nonumber \\
         h\sin\vartheta_1\sin\varphi_1 \!\!\!\!&=&\!\!\!\! \sin ^2 \theta_1 \dfrac{\varrho \Omega}{\Omega_{cr}}, \nonumber \\
         F_1 \sin(\vartheta_1 - \theta_1)\sin\varphi_1 \!\!\!\!&=&\!\!\!\! \sin\vartheta_1\bigg(\dfrac{\varrho \Omega}{\Omega_{r}} - h_{z}\bigg),
         \nonumber \\
         F_1 \sin\theta_1 \sin(\varphi_1 - \phi_1) \!\!\!\!&=&\!\!\!\! h \sin \varphi_1.
         \label{eq:res_rot_eq}
    \end{eqnarray}
Performing the direct integration of Eq.~(\ref{eq:def_Q}) and accounting the representation of the vectors $\mathbf{m}$ and $\mathbf{n}$ in the spherical coordinates
    \begin{eqnarray}
           &\dot{\mathbf{m}} = (-\varrho \Omega \sin\vartheta_1 \sin\Psi_1, \varrho \Omega \sin\vartheta_1 \cos\Psi_1, 0),\\
           &\mathbf{n} = (\sin\theta_1 \cos\Phi_1, \sin\theta_1 \sin\Phi_1, \cos\theta_1),
          \label{eq:m_n_spher_FA}
    \end{eqnarray}
where $\Psi_1 = \varrho \Omega t - \varphi_1$, we obtain
\begin{equation}
    \widetilde{Q} = 2 \varrho \widetilde{\Omega}\sin\vartheta_1\sin\varphi_1.
    \label{eq:Q_FA}
\end{equation}
Here, $\widetilde{\Omega} = \Omega/\Omega_{r1}$. It is important to note that Eq.~(\ref{eq:Q_FA}) is similar to the Eq.~(\ref{eq:Q_FP}) bearing in mind Eq.~(\ref{eq:FP_Phi}).

Using Eq.~(\ref{eq:def_Q}) and the representation of the vectors $\mathbf{m}$ and $\mathbf{n}$ in the spherical coordinates, one can derive the expression for the numerical calculation of the power loss. It corresponds to the expression (\ref{eq:Q_FM_Num}) with the following differences: $\widetilde{\tau} = \tau \Omega_{r1}$ (is chosen as $10^5$ in the simulations),
\begin{eqnarray}
    h_{xi} \!\!\!&=&\!\!\! h \cos(\varrho \widetilde{\Omega} \tilde{t}_i) + F_i \sin\theta_i\cos\phi_i,\nonumber
    \\
    h_{yi} \!\!\!&=&\!\!\! h \sin(\varrho \widetilde{\Omega} \tilde{t}_i) + F_i \sin\theta_i\sin\phi_i,\nonumber
    \\
    h_{zi} \!\!\!&=&\!\!\! h_z + F_i \cos\theta_i,
    \label{eq:h_FA_Num}
\end{eqnarray}
$F_i = F(\tilde{t}_i)$, $\theta_i = \theta(\tilde{t}_i)$, $\phi_i = \phi(\tilde{t}_i)$.

\begin{figure}
    \centering
    \includegraphics [width=1.0\linewidth] {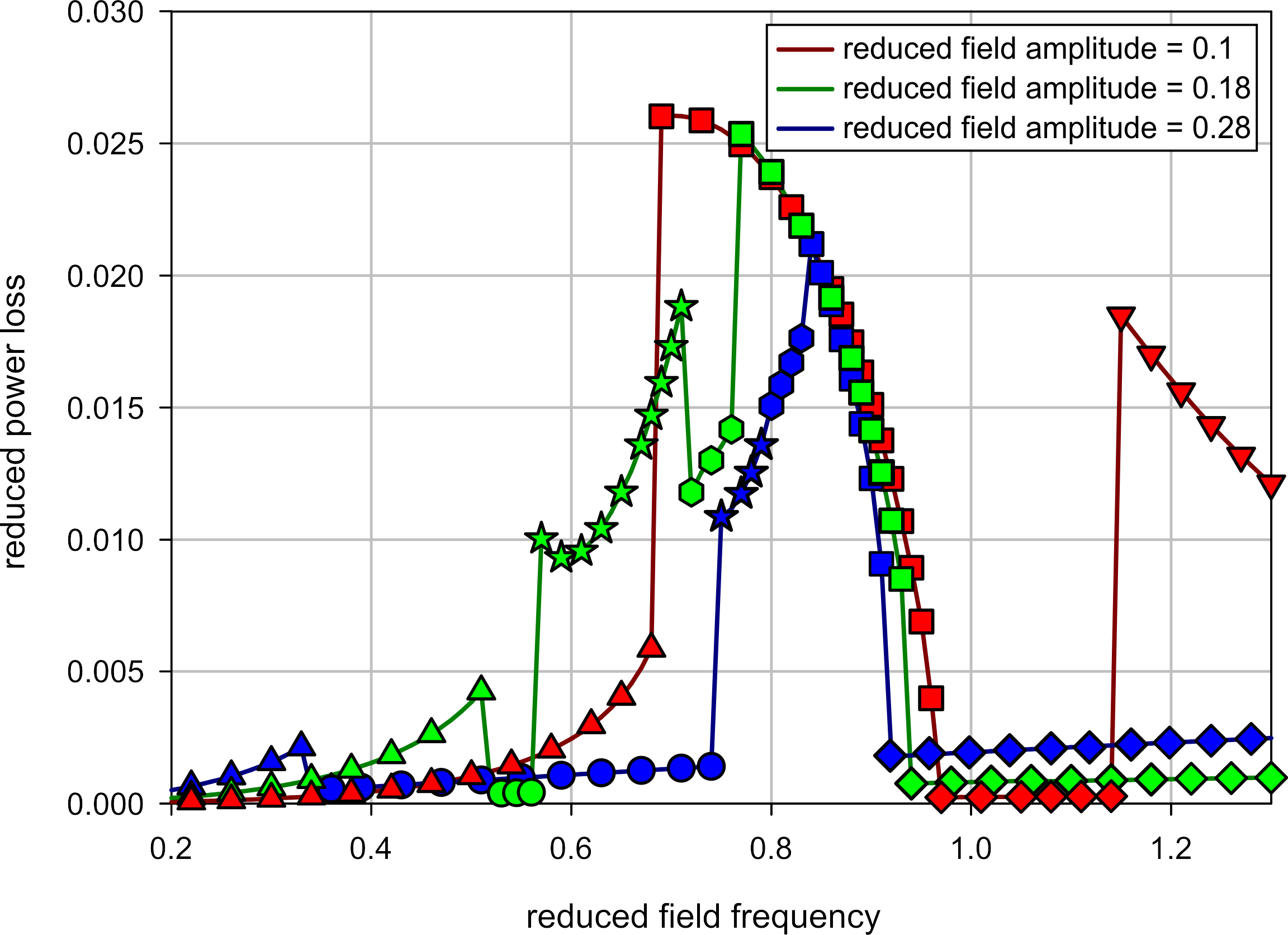}
        \caption {\label{fig:FA_Q_vs_w} (Color online) Model of the nanoparticle in a fluid with finite anisotropy: the most typical dependencies of the power loss on the field frequencies for different field amplitudes. The values of the system parameters are the following: $\alpha = 0.1$, $\eta = 0.006 P$, $M = 338 G$, $\varrho = +1$, $h_z = 0$. The simulation in the chosen range of the field parameters is not sensitive to the initial conditions. Triangle markers designate the uniform precession in the "up state"; circle markers designate the uniform precession after switching to the "down state"; star markers designate the nonuniform precession; square, hexagonal and down triangle markers designate the nonuniform mode with the immobilized magnetic moment in the "up state"; diamond markers designate the nonuniform mode with the immobilized magnetic moment in the "down state".}
\end{figure}

An additional degree of freedom inspires more interesting and complicated behaviour of the nanoparticle dynamics. First, like in the previous cases, a conventional nonuniform mode, which is characterised by oscillations of the precession angles, is generated. We recall that the period of these oscillations does not coincide with the field one. Second, like in the case of the fixed nanoparticle, the switching between two uniform modes, which is characterised by different nanoparticle orientations, can occur. Moreover, another one interesting mode takes place. It is characterised by the immobilized magnetic moment, while the nanoparticle body performs oscillations: the angles $\theta$ and $\phi$ of the vector $\mathbf{n}$ vary synchronously along with the external field without a recognizable drift, while the angles $\vartheta$ and $\varphi$ of the vector $\mathbf{m}$ remain practically constant. Such a type of motion is spread enough. It occurs in the "up state", "down state" and around the external field polarization plane. The transitions between the discussed types of motion are exhibited in the jumps in the dependence $\widetilde{Q}(\widetilde{\Omega}))$. Ultimately, we conclude that the switching between the uniform and nonuniform modes is accompanied by an abrupt increase in the power loss. The switching between two uniform modes with the nanoparticle reorientation leads, at least, to one-order reduction of the power loss. The most typical dependencies $\widetilde{Q}(\widetilde{\Omega})$ are depicted in Fig.~\ref{fig:FA_Q_vs_w}).

\section{Conclusions}
The forced coupled dynamics of the nanoparticle body and its magnetization has been considered in the deterministic approximation. The approach is based on the torque equation and equation of damped precession of the magnetization, which are derived from the total momentum conservation law. Within this framework, two modes of motion under the action of the circularly polarized field are described. While both the nanoparticle easy axis and its magnetic moment perform precession within the cones with constant angles, we call this the uniform mode. While the precession becomes unstable and the polar angle, at least, for the easy axis undergoes the periodical changes, we call this the nonuniform mode. To understand well the mechanisms of energy dissipation, we present the results for the coupled motion together with the results for the cases of the rigidly fixed nanoparticle inside the solid matrix and the rigidly fixed magnetization inside the mobile nanoparticle.

Since the external field is supposed to be rotating, it is natural that the simplest mode of the forced motion is the uniform precession. It is characterised by the constant precession angles and lag angles, the values of which depend on the system parameters. The algebraical equations for these dependencies have been obtained for all the cases considered. The main properties of the solution of the derived algebraical equations for the coupled dynamics case are the following. First, the nanoparticle magnetic moment always constitutes a smaller angle with the external field than with the anisotropy axis. Second, the precessional dynamics suggests the presence of some effective field, which is perpendicular to the field polarization plane, depends on the filed frequency and the polarization direction. For high frequencies, this effective field is large enough and can hold the magnetic moment almost along the direction, which is perpendicular to the polarization plane of the external field. The last fact was confirmed numerically. Finally, the expressions for the power loss have been obtained for the case of viscous rotation of the nanoparticle with finite anisotropy and for the simplified cases of the fixed nanoparticle and the fixed magnetization.

The nonuniform mode is a key issue of our investigations and has been described numerically. This mode consists in the periodical changes of the polar angles with a period, which does not coincide with the field one. The activation of the nonuniform precession mode is accompanied by the changes of the power loss. However, in the case of the motion of the nanoparticle with the magnetization fixed inside, the nonuniform precession leads to the power loss decrease. At the same time, in the case of the magnetic dynamics inside the fixed nanoparticle and in the case of the coupled motion of the nanoparticle with finite anisotropy in a viscous carrier, the nonuniform precession is connected with a considerable increase in the power loss. The coupled motion exhibits another type of the nonuniform precession. It is characterised by the nanoparticle magnetization, which is stable in the laboratory coordinates, and the nanoparticle body, which oscillates synchronously with the external field. This motion is realized for a wide enough range of parameters and can include several modes distinguished by discrete positions of the magnetization. The switching between them and the modes of other types is also connected with abrupt modifications of the power loss.

Despite the nonlinear effects in the coupled dynamics still need to be investigated thoroughly, the reported results allow us to state the following. The sharp transitions between the different precession modes is an important phenomenon to control the heating process within hyperthermia. On the one hand, it can be used to select the optimal parameters for therapy. On the other hand, we need to bear in mind it to prevent dangerous overheating. Although our approach does not take into account the thermal agitation and the dipole interaction between the nanoparticles, its relevance is obvious. First, as it follows from the approximation of the fixed magnetization, the interaction and thermal noise decrease these values. Therefore, the deterministic approach establishes the limit values of the power loss. Second, for the large enough nanoparticles ($\sim 20 \mathrm{nm}$) and comparatively intense external fields, the regular component in the nanoparticle dynamics is dominant. Consequently, the deterministic approach gives the results close to the correct ones.

\section*{Acknowledgment}
The authors express appreciation to V. V. Reva for the valuable help in the numerical simulations of the coupled magnetic and mechanical motion. Moreover, the authors are grateful to the Ministry of Education and Science of Ukraine for the financial support under Grant No. 0116U002622 and DAAD, the scholarship programme: Research Stays for University Academics and Scientists, 2018, Section: ST22, personal ref. no.: 91695699.

\section*{References}

\bibliography{Lyutyy_NAP_2018_Uniform_NonUniform_Dynamics}

\end{document}